\documentclass[12pt,journal,openany]{article}
\usepackage{graphicx}
\usepackage{amsmath}
\usepackage{soul}

\usepackage[top=30truemm,bottom=30truemm,left=25truemm,right=25truemm]{geometry}

\usepackage{amsmath,amssymb}
\usepackage{url}

\usepackage{array}
\usepackage{colortbl}
\definecolor{myCol}{rgb}{1,0.85,0.85}
\definecolor{myCol2}{rgb}{0.9,0.9,1}
\newcommand{\clcr}{\cellcolor{myCol}}
\usepackage{multirow}
\usepackage{caption}

\newcommand{\picWide}{3.3in}

\def\vec#1{ \mbox{\boldmath $#1$} }
\global\long\def\T#1{#1^{\top}}
\newcommand{\sumn}[1]{\sum_{#1=0}^{n-1}}
\newcommand{\prodn}[1]{\prod_{#1=0}^{n-1}}

\def\vec#1{ \mbox{\boldmath $#1$} }
\global\long\def\T#1{#1^{\top}}

\usepackage{xspace}
\newcommand{\etal}{{\it et al.}\xspace}

\newcommand{\DFTo}{${\rm DFTT}_{{\rm original}}$\xspace}
\newcommand{\DFTpre}{${\rm DFTT}_{{\rm present}}$\xspace}
\newcommand{\DFTpare}{${\rm DFTT}_{{\rm pareschi}}$\xspace}
\newcommand{\DFTpro}{${\rm DFTT}_{{\rm proposed}}$\xspace}
\newcommand{\Ho}{\mathcal{H}_0}
\newcommand{\II}{I\hspace{-.2em}I\xspace}

\usepackage{xcolor}

\renewcommand{\hl}[1]{#1}

\title{Randomness Evaluation with the Discrete Fourier Transform Test
Based on Exact Analysis of the Reference Distribution}
\author{Hiroki~Okada and~Ken~Umeno}

\begin{document}
	 \maketitle
\thanks{H. Okada and K. Umeno are with the Department
	 of Applied Mathematics and Physics, Graduate School of
	 Informatics, Kyoto University,
	 Kyoto, JAPAN.

	 e-mail: ir-okada@kddi.com, umeno.ken.8z@kyoto-u.ac.jp
	 }

\begin{abstract}
 In this paper, we study the problems in the \hl{discrete} Fourier
 \hl{transform} (DFT) test included in NIST SP 800-22 released by
 \hl{the} National Institute of
 Standards and Technology (NIST), which is a collection of tests for
 \hl{evaluating} both physical and \hl{pseudo-random} number generators for
 cryptographic applications. The most crucial problem in the DFT test is
 that its reference distribution of the test statistic is not derived
 mathematically \hl{but rather} numerically estimated; the DFT test for randomness
 is based on a pseudo-random number generator (PRNG). Therefore,
 the present DFT test should not be used unless the reference
 distribution is mathematically derived.
 Here, we prove that a power spectrum, which is a component of the test
 statistic, follows \hl{a} chi-squared distribution with 2 degrees of freedom.
 Based on this fact, we propose a test whose
 reference distribution of the test statistic is mathematically derived.
 Furthermore, \hl{the} results of testing non-random sequences and several PRNGs
 showed that the proposed test is more \textit{reliable}
 and definitely more \textit{sensitive} than the present DFT test.

\end{abstract}


 \section{Introduction}
 Random numbers are used in many \hl{types of applications}, such as
 cryptography, numerical \hl{simulations}, and so on. However, it is not easy to
 generate ``truly'' random number sequences. \hl{Pseudo-random}
 number generators (PRNGs) generate the sequences by iterating some recurrence
 relation\hl{;} therefore, the sequences are \hl{theoretically} not ``truly''
 random. The binary ``truly'' random sequence is defined as the
 sequence \hl{in which}
 each element has \hl{a} probability of exactly $\frac{1}{2}$ of being
 ``0'' or ``1'' and \hl{in which the elements} are statistically independent
 of each other. It is also
 difficult to ascertain if the sequence
 is truly random; \hl{therefore,} the randomness of the
 sequences is evaluated statistically.

  NIST SP 800-22 \cite{NIST_old,NIST_rev} is one of
 the famous statistical test
 suites for randomness that was used for \hl{selecting} the Advanced
 Encryption Standard (AES) algorithm.
 NIST SP 800-22
 consists of fifteen tests, and every test
 is hypothesis testing, \hl{where} the hypothesis is that \hl{the} input sequence is
 truly random; if the hypothesis is not rejected in all the tests,
 it is implied that the input sequences are random.
 Among the tests included in NIST SP 800-22, the DFT test is of \hl{the} greatest
 concern to us. This test detects  periodic features of a random number
 sequence; input sequences are \hl{discrete} Fourier \hl{transformed}, and the
 test statistic is  composed of the Fourier coefficients.
 In 2003, Kim \etal \cite{Kim_tec,Kim} reported that \hl{the} DFT test and the
 Lempel-Ziv test in the original NIST SP 800-22 \cite{NIST_old} have crucial
 theoretical problems. \hl{Regarding}
 the DFT test, it is reported that \hl{the} test statistic does not
 follow the expected reference distribution because of the problem
 that the DFT test regards Fourier coefficients as
 independent stochastic variables although they are not. Kim \etal
 numerically estimated the
 distribution of the test statistic with \hl{pseudo-random} numbers generated
 with a PRNG and proposed \hl{a}
 new DFT test with the estimated distribution.
 In 2005, Hamano \cite{Hamano} theoretically scrutinized the distribution of the
 Fourier coefficients in the original DFT test. However, he could not \hl{derive} the
 theoretical distribution of the test statistic, but \hl{he did make} the problems in
 the DFT test clearer.
 In 2005, because of these reports, in NIST SP 800-22 version 1.7,
 the Lempel-Ziv test was deleted, and the DFT test was revised
 according to the report of Kim \etal The DFT test has not
 \hl{subsequently been revised.}
 In 2012, Pareschi \etal \cite{Pareschi} reviewed three tests
 included in NIST SP 800-22, and they also numerically estimated the distribution of the
 test statistic. \hl{Consequently}, they reported that the distribution
 estimated by Kim \etal is not \hl{sufficiently accurate}.
 As stated above, several researchers have \hl{attempted} to revise the DFT test.
 However, the distribution of the test statistic \hl{has still} not been derived
 theoretically but \hl{rather} numerically estimated.

 In this paper, we review the problems in the DFT test, and \hl{we} prove three
 facts,  which are important \hl{for analyzing} the reference distribution of
 the test statistic: Under the assumption that the \hl{input}
 sequence is an ideal random number sequence, when $j\neq 0$,
 \begin{itemize}
  \item The asymptotic \hl{distributions} of both $\sqrt{\frac{2}{n}}c_j(X)$
	and $\sqrt{\frac{2}{n}}s_j(X)$ are the standard normal distribution
	($\mathcal{N}(0,1)$) when $n \to \infty$.

  \item When $n$ is sufficiently large, $\sqrt{\frac{2}{n}}c_j(X)$ and
	$\sqrt{\frac{2}{n}}s_j(X)$ are statistically independent \hl{of} each
	other.

  \item The asymptotic
	distribution of $\frac{2}{n}|S_j(X)|^2$
	is \hl{a} chi-squared distribution
	with 2 degrees of freedom $(\chi_2^2)$ when $n \to \infty$.

 \end{itemize}
 Here, $X$ is \hl{an} $n$-bit binary sequence, $S_j(X)$ is \hl{the} $j$-th
 discrete Fourier coefficient of $X$,  \hl{and} $c_j(X)$ and $s_j(X)$ are
 \hl{the} real and
 imaginary parts of $S_j(X)$, and they are defined in
 $\eqref{eq:defSj}$, $\eqref{eq:defcj} $ and $\eqref{eq:defsj}$ in
 Section $\rm{\ref{sec:DFT}}$, respectively.
 There is no information about these factors in NIST SP800-22, and,
 \hl{to the best of our knowledge}, no researchers who \hl{have studied} the DFT test have
 \hl{ever provided rigorous proofs}. These factors are necessary \hl{for analyzing} the
 reference distribution of the test statistic.
 Furthermore, we propose a new DFT test based on the \hl{fact} that
 $\chi_2^2$ is the asymptotic distribution of $\frac{2}{n}|S_j(X)|^2$.
 By comparing the results of several PRNGs, we show that
 our test is more \textit{reliable} and definitely \hl{more} \textit{sensitive}
 than the present DFT test.


 \section{Discrete Fourier Transform Test}
 \label{sec:DFT}
 In this section, we explain the procedure of the original DFT test
 (\DFTo),
 released in 2001 \cite{NIST_old}, before the revision in 2005
 \cite{NIST_rev}. We also explain the problems reported by several
 researchers \cite{Kim,Hamano}. The focus of this test is the peak
 heights in the \hl{discrete} Fourier \hl{transform} of the sequence. The purpose
 of this test is to detect periodic features in the tested sequence that
 would indicate a
 deviation from the assumption of randomness. The intention is to detect
 whether the number of peaks exceeding the 95 \% threshold is
 significantly different than 5 \%.


  \subsection{The procedure of the original DFT test}

  \label{subsec:Algorithm_DFT}
  \begin{enumerate}
   \item The zeros and ones of the input sequence  $E=\{\epsilon_0,\cdots,\epsilon_{n-1}\}$ are
	 converted to values of $-1$ and $+1$ to create the sequence
	 $X=\{x_0,\cdots,x_{n-1}\}$, where $x_i=2\epsilon_i-1 \ \ (i\in
	 \{0,\dots,n-1\})$. For simplicity, let $n$ be even.

   \item Apply a discrete Fourier transform (DFT) to $X$ to produce
	 Fourier coefficients $\{S_j(X)\}_{j=0}^{n-1}$.
	 The Fourier coefficient $S_j(X)$ and its real and imaginary
	 parts $c_j(X)$ and $s_j(X)$ are defined as follows:
	 \begin{small}
	  \begin{eqnarray}
	   S_j(X)&:=&\sum_{k=0}^{n-1}x_k\cos\frac{2\pi kj}{n}
	    -\sqrt{-1}\sum_{k=0}^{n-1}x_k\sin\frac{2\pi kj}{n} \label{eq:defSj} \\
	   c_j(X) &:=&\sum_{k=0}^{n-1}x_k\cos\frac{2\pi kj}{n}\label{eq:defcj}\\
	   s_j(X) &:=&\sum_{k=0}^{n-1}x_k\sin\frac{2\pi kj}{n}\label{eq:defsj}
	  \end{eqnarray}
	 \end{small}

   \item Compute $\{|S_j(X)|\}_{j=0}^{\frac{n}{2}-1}$, where
	 \[
	 |S_j(X)|^2=(c_j(X))^2 +(s_j(X))^2.
	 \]
	 Because $|S_j(X)|=|\overline{S_{n-j}(X)}|$,
	 $\{|S_j(X)|\}_{j=\frac{n}{2}}^{n-1}$ are discarded.

	 \label{enum:Zj}

   \item Compute a threshold value $T_{0.95}=\sqrt{3n}$.
	 \hl{The} 95\% values
	 $\{|S_j(X)|\}_{j=0}^{\frac{n}{2}-1}$ are supposed to be
	 $<T_{0.95}$.

	 According to SP800-22, $\frac{2}{n}|S_j(X)|^2$ is considered to
	 follow $\chi_2^2$, and $T_{0.95}$ is defined by \hl{the} following
	 equation.
	 \begin{eqnarray*}
	  P(|S_j(X)|<T_{0.95})
	   &=&\int_0^{\frac{2}{n} T_{0.95}^2}
	   \frac{1}{2}e^{-\frac{y}{2}} dy\\
	   &=&
	   1-e^{-\frac{T_{0.95}^2}{n}}\\
	  &:=&0.95\\
	  \therefore
	   T_{0.95} &=& \sqrt{-n\ln(0.05)} \simeq \sqrt{3n}
	 \end{eqnarray*}

	 Several researchers \cite{Kim,Hamano} reported
	 \hl{that} this $T_{0.95}=\sqrt{3n}$ was incorrect, and it was accordingly
	 revised as $T_{0.95}=\sqrt{-n\ln(0.05)}$ in the DFT
	 test in the revised NIST SP800-22 \cite{NIST_rev}.

   \item Count
	 \[
	 N_1 = \# \left\{|S_j(X)| \ | \ |S_j(X)| < T_{0.95}, 0 \le j \le \frac{n}{2}-1\right\}.
	 \]
	 If $\{|S_j(X)|\}^{\frac{n}{2}-1}_{j=0}$ \hl{are} mutually
	 independent, then under the assumption of randomness,
	 $N_1$ can be considered to follow
	 $\mathcal{B}(\frac{n}{2},0.95)$, where $\mathcal{B}$ is the
	 binomial distribution.

	 According to the central limit theorem, when $n$ is
	 \hl{sufficiently large}, the approximation to $\mathcal{B}(n,p)$ is given
	 by the normal distribution $\mathcal{N}(np,\,np(1-p))$.
	 Therefore, when $n$ is sufficiently large, under the assumption
	 of randomness,
	 \[
	 N_1 \sim
	 \mathcal{N}\left(0.95\frac{n}{2},(0.95)(0.05)\frac{n}{2}\right).
	 \]

	 \label{enum:binom}

   \item Compute a test static
	 \[
	 d=\frac{N_1-0.95 \frac{n}{2}}{\sqrt{(0.95)(0.05)\frac{n}{2}}}.
	 \]
	 When $n$ is sufficiently large, under the assumption of randomness, the test
	 statistic $d$ can be considered to follow $\mathcal{N}(0,1)$ \label{enum:d}

   \item Compute $P$-$value$; ${\displaystyle p ={\rm erfc}\left(\frac{|d|}{\sqrt{2}}\right)}$.

	 If  $p < \alpha$, then conclude that
	 the sequence is non-random, where
	 $\alpha$ is a significance level of the DFT test. NIST
	 recommends $\alpha=0.01$ \cite{NIST_rev}. Therefore, we also define
	 $\alpha=0.01$.
	 If  $p \ge \alpha$, conclude that the
	 sequence is random.

   \item \hl{Perform} 1) to 7) for $m$ sample sequences $\{X_1,X_2,\dots, X_m\}$;
	  $m$ $P$-$value$s $\{p_1,p_2,\dots,p_m\}$ are computed.

   \item (Second-level test I: Proportion of sequences passing a test)

	 Count the number of sample sequences for which
	 $P$-$value$ $\ge \alpha$ and define it as $m_{p}$.
	 Then, under the assumption of randomness, $m_p$ follows
	 $\mathcal{B}(m,1-\alpha)$, which approximates
	 $\mathcal{N}(m(1-\alpha),m\alpha(1-\alpha))$ when $m$ is
	 \hl{sufficiently large}. Therefore,
	 the proportion of sequences passing a test ($= m_p/m$)
	 approximately follows
	 $\mathcal{N}\left((1-\alpha),\frac{\alpha(1-\alpha)}{m}\right)$.
	 The range of acceptable $m_p/m$ is determined using the
	 significance interval defined as
	 \begin{small}
	  \begin{eqnarray}
	   1 - \alpha - 3
	    \sqrt{\frac{\alpha (1 - \alpha)}{m}} < \frac{m_p}{m}
	    <  1 - \alpha + 3
	    \sqrt{\frac{\alpha (1 - \alpha)}{m}}. \label{eq:sig_interval}
	  \end{eqnarray}
	 \end{small}
	 If the proportion falls outside of this
	 interval, there is evidence that the data are
	 non-random.
	 \label{enum:Prop_Pass}

   \item (Second-level test II: Uniform distribution of $P$-$value$s)

	 Uniformity may also be determined by applying a $\chi^2$
	 test and determining a $P$-$value$ corresponding to the
	 goodness-of-fit distributional test on the $P$-$value$s obtained
	 for an arbitrary statistical test (i.e., the $P$-$value$ of the
	 $P$-$value$s). This is \hl{performed} by computing
	 \begin{eqnarray*}
	  \chi^2&=&\sum_{i=1}^{10} \frac{(F_i-m/10)^2}{m/10},
	 \end{eqnarray*}
	 where $F_i$ is the number of $P$-$value$s
	 in sub-interval $i$. A $P$-$value$ $P_T$
	 is calculated such that
	 \[
	 P_{T} = \rm{igamc}\left(\frac{9}{2} ,\frac{\chi^2}{2}\right),
	 \]
	 where igamc is the complementary incomplete gamma function.
	 If
	 \begin{eqnarray}
	  P_T \ge \alpha_{\II} (:=0.0001), \label{eq:alpha_II}
	 \end{eqnarray}
	 the sequences can be considered to be
	 uniformly distributed, where $\alpha_{\II}$ is the significance
	 level for $P_T$.
	 \label{enum:Uniformity_test}

   \item  If the set of $P$-$value$s $\{p_1,p_2,\dots,p_m\}$ passes
	  both \ref{enum:Prop_Pass}) and \ref{enum:Uniformity_test}),
	  the physical or \hl{pseudo-random} number generators that generated
	  \hl{the input sequences are} concluded to be ideal.
  \end{enumerate}


  \subsection{The fundamental problems of the original and present DFT \hl{tests}}
  Kim \etal \cite{Kim} and Hamano \cite{Hamano} reported the following:
  \begin{itemize}
   \item The test statistic $d:=\frac{N_1-0.95
	 \frac{n}{2}}{\sqrt{(0.95)(0.05)\frac{n}{2}}}$ does not follow
	 $\mathcal{N}(0,1)$;

   \item $N_1$ does not follow
       $\mathcal{N}\left(0.95\frac{n}{2},(0.95)(0.05)\frac{n}{2}\right)$.
  \end{itemize}
  Furthermore, Kim \etal, using Secure Hash Generator
  (G-SHA1) \cite{NIST_rev} as a PRNG, estimated that
  \begin{eqnarray*}
   N_1 &\sim&  \mathcal{N}\left(0.95\frac{n}{2},(0.95)(0.05)\frac{n}{4}\right);\\
   d_{kim} &:=& \frac{N_1-0.95\frac{n}{2}}{\sqrt{(0.95)(0.05)\frac{n}{4}}}
    \sim \mathcal{N}(0,1),
  \end{eqnarray*}
  and \DFTo was revised according to this report of Kim  \etal
  \cite{NIST_rev}; the present DFT test, denoted as \DFTpre,
  has not been revised since then.
  Therefore, the reference distribution of the test statistic of
  \DFTpre is not mathematically derived.
  Furthermore, Pareschi \etal reported that the numerical
  estimation is {\it not} \hl{sufficiently accurate}; they numerically estimated that
  \begin{eqnarray*}
   N_1 &\sim&  \mathcal{N}\left(0.95\frac{n}{2},(0.95)(0.05)\frac{n}{3.8}\right);\\
   d_{pareschi} &:=& \frac{N_1-0.95\frac{n}{2}}{\sqrt{(0.95)(0.05)\frac{n}{3.8}}}
    \sim \mathcal{N}(0,1).
  \end{eqnarray*}
  \hl{Moreover}, Pareschi \etal proposed that the DFT test with this test statistic
  (\DFTpare) is more \textit{reliable}. (\hl{The definition of
  the} \textit{reliability} of a test is discussed in Section \ref{sec:Experiment_DFT}.)
  Therefore, it can be \hl{considered} that \DFTpre still has errors.
  \hl{First}, \DFTpre and \DFTpare are \hl{performed} based on a PRNG, whose randomness should
  be evaluated with a randomness test; they cannot be used unless the
  reference distribution is mathematically derived.

  As stated in step 5) in Section \ref{subsec:Algorithm_DFT},
  $\{|S_j(X)|\}^{\frac{n}{2}-1}_{j=0}$ are considered to be mutually
  independent. However, $\{|S_j(X)|\}^{\frac{n}{2}-1}_{j=0}$
  are not mutually independent, and this problem is expected to be the
  main factor for \hl{why} $N_1$ does not follow
  $\mathcal{N}\left(0.95\frac{n}{2},(0.95)(0.05)\frac{n}{2}\right)$
  \cite{Kim,Hamano}.
  Furthermore, before \hl{considering} this
  problem, it is also \hl{necessary to ensure} that $\frac{2}{n}|S_j(X)|^2$ follows
  $\chi_2^2$.  Although $\frac{2}{n}|S_j(X)|^2$ is considered to \hl{follow}
  $\chi_2^2$ in step 4) in Section \ref{subsec:Algorithm_DFT}, there is
  no information about this in SP800-22, and no
  researchers studying
 the DFT test have \hl{ever provided rigorous proofs to the best of our
  knowledge. We provide a proof
  for the DFT test in Section} \ref{sec:Z_j}.


 \section{The asymptotic distribution of $\frac{2}{n}|S_j(X)|^2$}
 \label{sec:Z_j}
  In this section, we analyze the asymptotic distribution of
  $\frac{2}{n}|S_j(X)|^2$. From the definition of $|S_j(X)|$ in \eqref{eq:defSj},
  \begin{eqnarray*}
   \frac{2}{n}|S_j(X)|^2
    &=&\left(\sqrt{\frac{2}{n}}c_j(X)\right)^2
    +\left(\sqrt{\frac{2}{n}}s_j(X)\right)^2.
  \end{eqnarray*}
  When $j=0$,
  \[
  \frac{2}{n}|S_0(X)|^2=2\left(\frac{\sum_{k=0}^{n-1}x_k}{\sqrt{n}}\right)^2.
  \]
  Under the assumption that $X$ is an ideal random number sequence,
  $P(x_k=-1)=P(x_k=1)=\frac{1}{2}$ and $\{x_k\}_{k=0}^{n-1}$ are
  mutually independent, and $E[x_k]=0,V[x_k]=1$.
  Therefore,
  as a consequence of the central limit theorem,
  when $n$ is \hl{sufficiently large},
  $\left(\frac{\sum_{k=0}^{n-1}x_k}{\sqrt{n}}\right)$ follows $\mathcal{N}(0,1)$,
  and $\left(\frac{\sum_{k=0}^{n-1}x_k}{\sqrt{n}}\right)^2$ follows
  \hl{a} chi-squared distribution
  with 1 \hl{degree} of freedom $(\chi_1^2)$.
  Thus, $\frac{2}{n}|S_0(X)|^2$ does not \hl{follow} $\chi_2^2$.

  In the following, we consider the case when $j\neq 0$.
  Here,  $\frac{2}{n}|S_j(X)|^2$ follows $\chi_2^2$ if the following is true:
  \begin{itemize}
   \item Both $\sqrt{\frac{2}{n}}c_j(X)$ and  $\sqrt{\frac{2}{n}}s_j(X)$
	 \hl{follow} $\mathcal{N}(0,1)$.
   \item $\sqrt{\frac{2}{n}}c_j(X)$ and  $\sqrt{\frac{2}{n}}s_j(X)$ are
	 mutually independent.
  \end{itemize}
  In the following 2 subsections, we prove the following
  Theorem 1, Theorem 2 and Theorem 3:
  \vspace{-3mm}
  \begin{flushleft}
   \begin{tabular}{lp{6.5cm}}
    Theorem 1: & When $n$ is sufficiently large, both $\sqrt{\frac{2}{n}}c_j(X)$
    and  $\sqrt{\frac{2}{n}}s_j(X)$ follow $\mathcal{N}(0,1)$.  \\
    Theorem 2:& When $n$ is sufficiently large, $\sqrt{\frac{2}{n}}c_j(X)$ and
	$\sqrt{\frac{2}{n}}s_j(X)$ are mutually independent.\\
    Theorem 3:& $\frac{2}{n}|S_j(X)|^2$ follows $\chi_2^2$ when $n$ is sufficiently large.
   \end{tabular}
  \end{flushleft}
   \noindent From the definition of $\chi_2^2$, Theorem 3 can be
   proven by combing Theorem 1 and Theorem 2.

  \subsection{Proof of Theorem 1: The asymptotic distribution of $\sqrt{\frac{2}{n}}c_j(X)$}
  In this subsection, we prove Theorem 1.
  Hamano \cite{Hamano} showed that \hl{the} average, variance, skewness,
  \hl{and} kurtosis
  of $c_j(X)$ and $\mathcal{N}(0,\frac{n}{2})$ are \hl{the} same. However, it
  \hl{cannot} be proven that $\mathcal{N}(0,\frac{n}{2})$ is the asymptotic
  distribution of $c_j(X)$ based only on these factors.

  $\sqrt{\frac{2}{n}}c_j(X)$ is expressed as  $\sqrt{\frac{2}{n}} c_j(X)
  :=\sqrt{\frac{2}{n}}\sum_{k=0}^{n-1}x_ka_{k,j}$,
  where $a_{k,j}=\cos\frac{2\pi kj}{n}$. Under the assumption that $X$
  is an ideal random number sequence, the  characteristic function of
  $\sqrt{\frac{2}{n}}c_j(X)$ denoted by $\phi(t)$ is expressed as \hl{follows}:
  \begin{eqnarray*}
   \phi(t) &=& E_X\left[\exp\left(\sqrt{\frac{2}{n}}\sqrt{-1} tc_j(X)\right)\right]\\
   &=& E_X\left[\prod_{k=0}^{n-1}\exp\left(\sqrt{\frac{2}{n}}\sqrt{-1}t x_k
				   a_{k,j}\right)\right]\\
   &= &
    \prod_{k=0}^{n-1}E_{x_k}\left[\exp\left(\sqrt{\frac{2}{n}}\sqrt{-1}t x_k
				       a_{k,j}\right)\right]\\
   &=&
    \prod_{k=0}^{n-1}
    \cos\left(\sqrt{\frac{2}{n}}t a_{k,j}\right).\\
   \therefore  \log\phi(t)&=&\sum_{k=0}^{n-1} \log\cos\left(\sqrt{\frac{2}{n}}t
  a_{k,j}\right),
  \end{eqnarray*}
  where
  \[
  E_X(\cdot):=\frac{1}{2^n}\sum_{X \in \{-1,1\}^n}(\cdot),
  \]
  \[
  E_{x_k}(\cdot):=\frac{1}{2}\sum_{x_k \in \{-1,1\}}(\cdot).
  \]
  Using the Taylor expansion about a point $t=0$, we
  obtain
  \[
    \log\cos\left(\sqrt{\frac{2}{n}}t a_{k,j}\right)
     =
     -\frac{1}{n}a_{k,j}^2t^2
     -\frac{1}{3n^2}a_{k,j}^4t^4
     +O(t^6).
  \]
  \[
   \therefore \log\phi(t)
    =
     -\frac{1}{n}\sumn{k}a_{k,j}^2t^2
     -\frac{1}{3n^2}\sumn{k}a_{k,j}^4t^4
     +O(t^6).
  \]
  Since
  \[
  \sum_{k=0}^{n-1} a_{k,j}^2=\frac{n}{2}, \,\,
  \sum_{k=0}^{n-1} a_{k,j}^{2l}\le n  \  \ (l \in \{1,2,3,\dots\}),
  \]

  \begin{eqnarray*}
   \lim_{n\to\infty}\log\phi(t)=-\frac{1}{2}t^2.
     \ \ \therefore
    \lim_{n\to\infty}\phi(t)=e^{-\frac{1}{2}t^2}.
  \end{eqnarray*}
  Thus, $\mathcal{N}(0,1)$ is the asymptotic distribution of
  $\sqrt{\frac{2}{n}}c_j(X)$. Likewise, it can be proven that
  $\mathcal{N}(0,1)$ is the asymptotic distribution of
  $\sqrt{\frac{2}{n}}s_j(X)$.


  \subsection{Proof of Theorem 2: Statistical independence of $\sqrt{\frac{2}{n}}c_j(X)$  and $\sqrt{\frac{2}{n}}s_j(X)$}
  In this subsection, we prove Theorem 2.
   Let us define a 2-dimensional stochastic
   variable $\vec{Y}$ as \hl{the} following equation:
  \begin{eqnarray*}
   \vec{Y} &:= & (Y_1,Y_2):=\left(\sqrt{\frac{2}{n}}c_j(X),\sqrt{\frac{2}{n}}s_j(X)\right).
  \end{eqnarray*}
  Under the assumption that $X$ is an ideal random number sequence,
  \hl{the} characteristic function of
  $\vec{Y}$ denoted by $\psi(\vec{t})$ is expressed as \hl{follows}:
  \begin{eqnarray*}
   \psi(\vec{t})&=&E_X[\exp(\sqrt{-1}\vec{t}\T{\vec{Y}})]\\
   &=& E_{X}\left[\exp\left(\sqrt{\frac{2}{n}}\sqrt{-1}(t_1c_j(X)+t_2s_j(X))\right)\right]\\
   &= &
    \prodn{k}
    \cos\left(\sqrt{\frac{2}{n}}
	 \left(
	  t_1a_{k,j}
	  +t_2b_{k,j}
	 \right)
	\right),
\end{eqnarray*}
  where
  \[
  \vec{t}=(t_1,t_2), \, a_{k,j}=\cos\frac{2\pi kj}{n}, \,
  b_{k,j}=\sin\frac{2\pi kj}{n}.
  \]
  Therefore,
  \begin{eqnarray*}
   \log\psi(\vec{t})=
    \sumn{k}
    \log\cos\left(\sqrt{\frac{2}{n}}
	     (a_{k,j}t_1+b_{k,j}t_2)
	    \right).
  \end{eqnarray*}
  Using the Taylor expansion about a point $\vec{t}=\vec{0}$, we obtain
  \begin{eqnarray*}
   &&
    \log\cos\left(\sqrt{\frac{2}{n}}(a_kt_1+b_kt_2)
	    \right)\\
   &=&
    -\frac{(a_{k,j}t_1+b_{k,j}t_2)^2}{n}
    -\frac{(a_{k,j}t_1+b_{k,j}t_2)^4}
    {3n^2}+\cdots.
  \end{eqnarray*}
Since
\begin{eqnarray*}
\sumn{k}a_k^2=\sumn{k}b_k^2=\frac{n}{2},\,\,
\sumn{k}a_kb_k = 0,\,\,
\sumn{k}a_k^lb_k^m \le n \, (l,m\ge0),\,\,
\end{eqnarray*}
we obtain
\begin{eqnarray*}
 \lim_{n\to\infty}\log\psi(\vec{t})
  = -\frac{\vec{t}\T{\vec{t}}}{2},  \ \ \ \therefore
 \lim_{n\to\infty}\psi(\vec{t}) = \exp\left(-\frac{\vec{t}\T{\vec{t}}}{2}\right).
\end{eqnarray*}
Therefore, when $n$ is sufficiently large, the joint probability
distribution function is described as \hl{follows}:
\begin{eqnarray*}
f_{Y_1,Y_2}(y_1,y_2)
&=& \frac{1}{2\pi}\exp\left(-\frac{y_1^2+y_2^2}{2}\right).
\end{eqnarray*}
As we proved before, $\mathcal{N}(0,1)$ is the asymptotic distribution
of both $Y_1$ and $Y_2$. Thus, when $n$ is sufficiently large,
the probability distribution \hl{functions} of $Y_1$ and $Y_2$ \hl{are}
$f_{Y_1}(y_1)= \frac{1}{\sqrt{2\pi}}\exp\left(-\frac{y_1^2}{2}\right)$
and $f_{Y_2}(y_2)=
\frac{1}{\sqrt{2\pi}}\exp\left(-\frac{y_2^2}{2}\right)$, respectively.
Therefore, when $n$ is sufficiently large, the following
equation is obtained:
\begin{eqnarray*}
 f_{Y_1,Y_2}(y_1,y_2)= f_{Y_1}(y_1)f_{Y_2}(y_2).
\end{eqnarray*}
This means that $\sqrt{\frac{2}{n}}c_j(X)$ and
$\sqrt{\frac{2}{n}}s_j(X)$ are \textit{mutually independent} when $n$ is
sufficiently large.


 \section{The proposed DFT test}

 In  Section \ref{sec:Z_j}, we proved Theorem 3, stating that
$\frac{2}{n}|S_j(X)|^2
 (j\neq 0)$
 follows $\chi_2^2$ when $n$ is sufficiently large. Therefore, if
 $\{|S_j(X)|\}^{\frac{n}{2}-1}_{j=1}$ \hl{are}  mutually independent,  we
 can consider that $N_1$ follows $\mathcal{N}\left(0.95\frac{n}{2},
 (0.95)(0.05) \frac{n}{2}\right)$. However,
 $\{|S_j(X)|\}^{\frac{n}{2}-1}_{j=1}$ are not mutually independent.
 Therefore, it is \hl{necessary} to  mathematically analyze the distribution
 of the test statistic $d$ under the condition that
 $\{|S_j(X)|\}^{\frac{n}{2}-1}_{j=1}$ are not mutually independent.
 Hamano \cite{Hamano} \hl{attempted} to mathematically derive the distribution of
 the set $\{|S_j(X)|\}^{\frac{n}{2}-1}_{j=1}$, but he could not do \hl{so},
 and we also could not derive \hl{this distribution}. However, we rigorously proved that the
 asymptotic distribution of $\frac{2}{n}|S_j(X)|^2$ is $\chi_2^2$, and
 \hl{we develop} the new
 DFT test (\DFTpro)  based on this fact.
 The reference distribution of the test statistic of \DFTpro is
 mathematically derived, \hl{whereas}
 that of \DFTpre \hl{is} estimated with a PRNG. We explain the test statistic of
 \DFTpro in the next subsection.


  \subsection{The procedure of the proposed DFT test}
  \label{sec:procedure_DFTpro}
  \hl{In the standard approach in NIST SP800-22, each sequence is analyzed; thus, $m$
  sequences give $m$ $P$-$value$s.
  However, } \DFTpro \hl{generates $\frac{n}{2}-1$
  ($n$: length of a sequence) $P$-$value$s.
  Therefore, more  $P$-$value$s are generated since $n$ is generally
  larger than $m$. Since the number of $P$-$value$s should not be too
  large (see Section $\ref{sec:parameter}$),
  before conducting }\DFTpro, \hl{it is necessary to adjust the length of the
  sequences and make them into more sets of short sequences (see also
  Table~$\ref{tab:ParameterNM}$),
  assuming that the set input sequences are continuously generated by an RNG.
  Therefore, }\DFTpro \hl{is theoretically not appropriate for the isolated set
  of sequences.}

  \hl{The procedure of the proposed DFT test is described as follows}:
  \begin{enumerate}
   \item The zeros and ones of the $m$ $n$-length input sequence
	 $\{E_i=\{\epsilon_0^i,\cdots,\epsilon_{n-1}^i\}\}_{i=1}^m$ are
	 converted to values of $-1$ and $+1$ to create the sequence
	 $\{X^i=\{x^i_0,\cdots,x^i_{n-1}\}\}_{i=0}^m$, where $x^i_j=2\epsilon_j-1 \ \ (j\in
	 \{0,\dots,n-1\})$. For simplicity, let $n$ be even.

   \item Apply a discrete Fourier transform (DFT) to each $X^i$ to produce
	 Fourier coefficients $\{S_j(X^i)\}_{j=0}^{n-1}$.
	 The Fourier coefficient $S_j(X^i)$ and its real and imaginary
	 parts $c_j(X^i)$ and $s_j(X^i)$ are \hl{defined as follows:}
	 \begin{eqnarray*}
	  S_j(X^i)&:=&\sum_{k=0}^{n-1}x_k\cos\frac{2\pi kj}{n}
	   -\sqrt{-1}\sum_{k=0}^{n-1}x_k\sin\frac{2\pi kj}{n}, \\
	  c_j(X^i) &:=&\sum_{k=0}^{n-1}x_k\cos\frac{2\pi kj}{n},\\
	  s_j(X^i) &:=&\sum_{k=0}^{n-1}x_k\sin\frac{2\pi kj}{n},
	 \end{eqnarray*}

   \item For all $j \in \{1,\dots,\frac{n}{2} - 1\}$, perform the
	 Kolmogorov-Smirnov (KS)
	 test \cite{Stephen,Thode} on the
	 empirical cumulative distribution function of
	 $\{\frac{2}{n}S_j(X_i)\}_{i=1}^m$ defined as $F_{m}^j(y)$ based on the
	 difference from $\chi_2^2$ and compute the $P$-$value$ $p_j$. Here,
	 the KS statistic $D_{m}^j$ and $p_j$ are defined as follows.
	 \begin{eqnarray*}
	  D_{m}^j&=&\sqrt{m}\max_{y>0}\left|F_{m}^j(y)-F(y)\right|,\\
	  p_j    &=&1-H(D_{m}^j),
	 \end{eqnarray*}
	 where $H(y)$ is the cumulative distribution function of
	 the Kolmogorov-Smirnov distribution:
	 \[
	 H(y)=1-2\sum_{i=1}^\infty(-1)^{i-1}e^{-2i^2y}.
	 \]
	 Note that $\frac{n}{2}-1$ $P$-$value$s $\{p_1,p_2,\dots,p_{\frac{n}{2}-1}\}$
	 are computed in this step, while the \DFTpre \hl{computes}
	 $m$ $P$-$value$s.
	 \label{sec:Pvalue_DFTpro}

   \item Perform the \hl{second-level tests} I and II
	 defined in the original DFT test (\hl{see} Section
	 $\ref{subsec:Algorithm_DFT}$-$\ref{enum:Prop_Pass}$,
	 $\ref{subsec:Algorithm_DFT}$-$\ref{enum:Uniformity_test}$).
	 If the set of $P$-$value$s $\{p_1,p_2,\dots,p_{\frac{n}{2}-1}\}$ passes
	 both \hl{second-level tests} I and II, the physical or pseudo-random
	 number generator that generated \hl{the} input sequences is concluded to be ideal.

  \end{enumerate}


 \section{Experiments}
 \label{sec:Experiment_DFT}
 In this section, we explain the experiments \hl{that we performed and the conclusions}
 derived from their results. In these experiments, we compare the
 \textit{reliability} and \textit{sensitivity} of \DFTpre
 and \DFTpro.
\hl{The} \textit{reliability} of tests means \hl{a} low probability
 of \textit{false \hl{positives}} (type I error) (\hl{see} Table \ref{tab:TypeofError}), and
\hl{the}  \textit{sensitivity} of tests means \hl{a} low probability
 of \textit{false \hl{negatives}} (type II error). Now, the null hypothesis of the
 tests ($\mathcal{H}_0$) is that \hl{the} ``generator is ideal''. Therefore,
\hl{a}  \textit{false positive} (type I error) means an erroneous
\hl{identification} of an
 ideal generator as not random,
 and \hl{a} \textit{false negative} (type II error) means an erroneous
 \hl{identification of a
 generator that} is not ideal as random.
 Comparing the probability of type I error and type II error, we can
 conclude which test is better.
 \begin{table}[!t]
  \centering
  \caption{Types of error}
  \label{tab:TypeofError}
  \begin{tabular}{|c|c|c|c|}
   \hline
   \multicolumn{2}{|c|}{$\mathcal{H}_0:$ Null hypothesis}&
   \multicolumn{2}{c|}{$\mathcal{H}_0$ is}\\
   \cline{3-4}
   \multicolumn{2}{|c|}{= ``generator is ideal''}
   & True  &   False \\
   \hline
   \multirow{4}{*}{Judgment of $\mathcal{H}_0$}
   &\multirow{2}{*}{Reject} & False Positive  & \multirow{2}{*}{True Positive} \\
   &  &  (Type I error)  &   \\
   \cline{2-4}
   &\multirow{2}{*}{Fail to reject} &
   \multirow{2}{*}{True Negative}  & False Negative \\
   &  &    & (Type II error)  \\
   \cline{2-4}
   \hline
  \end{tabular}
 \end{table}

For simplicity, in this experiment, we modify \hl{the} significance interval of
the second-level test I defined
in \eqref{eq:sig_interval} as follows:
\begin{small}
 \begin{eqnarray}
  1 - \alpha - 2.575
   \sqrt{\frac{\alpha (1 - \alpha)}{m}} < \frac{m_p}{m}
   <  1 - \alpha + 2.575
   \sqrt{\frac{\alpha (1 - \alpha)}{m}}. \label{eq:sig_interval2}
 \end{eqnarray}
\end{small}
 With this modified significance interval, the significance level of \hl{the}
 second-level test I ($:=\alpha_I$) is modified to be $\alpha_I=0.01$.


  \subsection{Experiment 1: Test results for periodic sequences}
  \begin{figure}[!t]
   \centering
   \includegraphics[width=\picWide,bb=0 0 1200 1500]{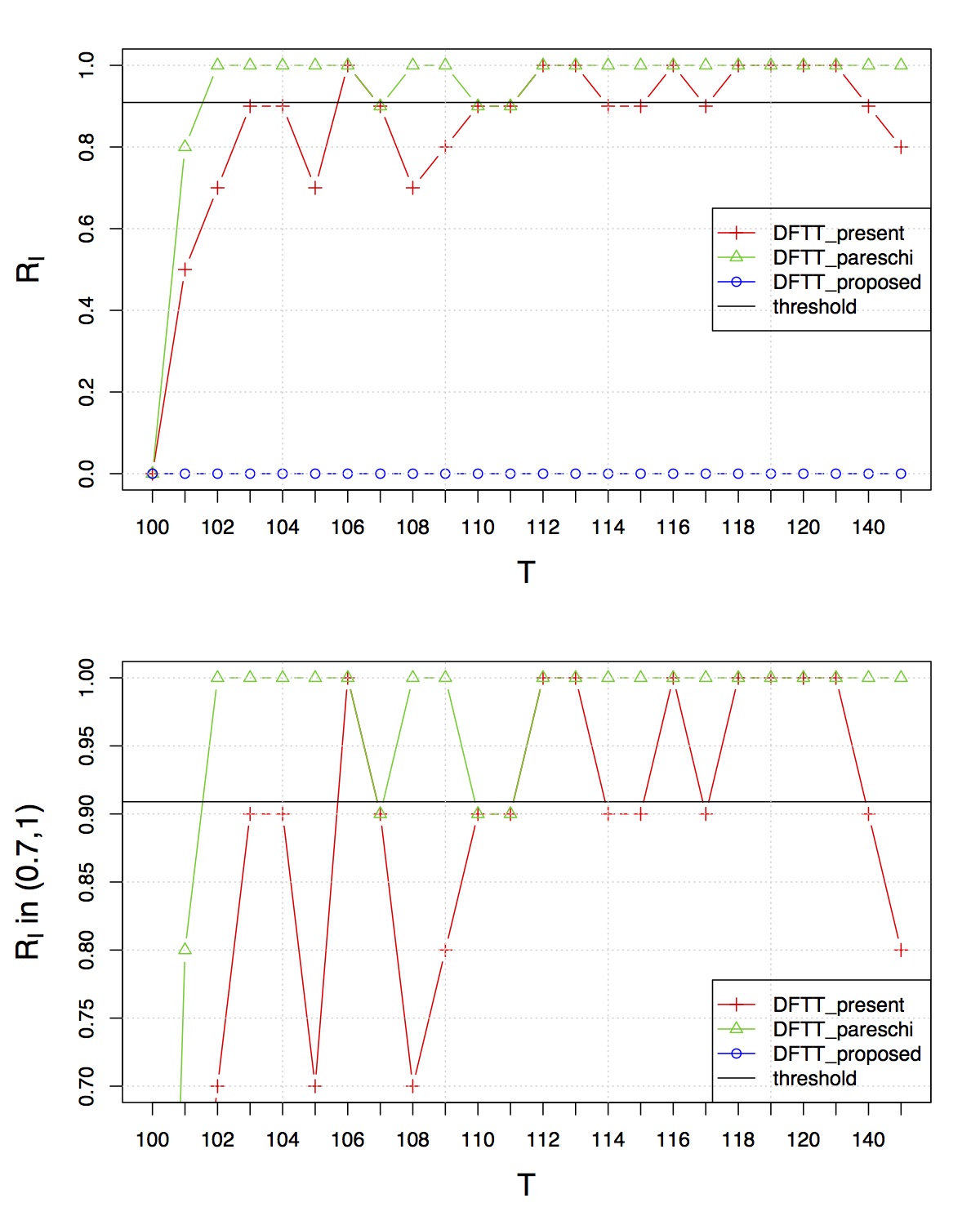}
   \caption{Passing rate $R_I$ in experiment 1.
   The ``threshold'' means the lower limit of the significance
   interval defined in Eq. \eqref{eq:sig_ex1_I}}
   \label{fig:R_I_T}
  \end{figure}

  \begin{figure}[t]
   \centering
   \includegraphics[width=\picWide,bb=0 0 1200 1500]{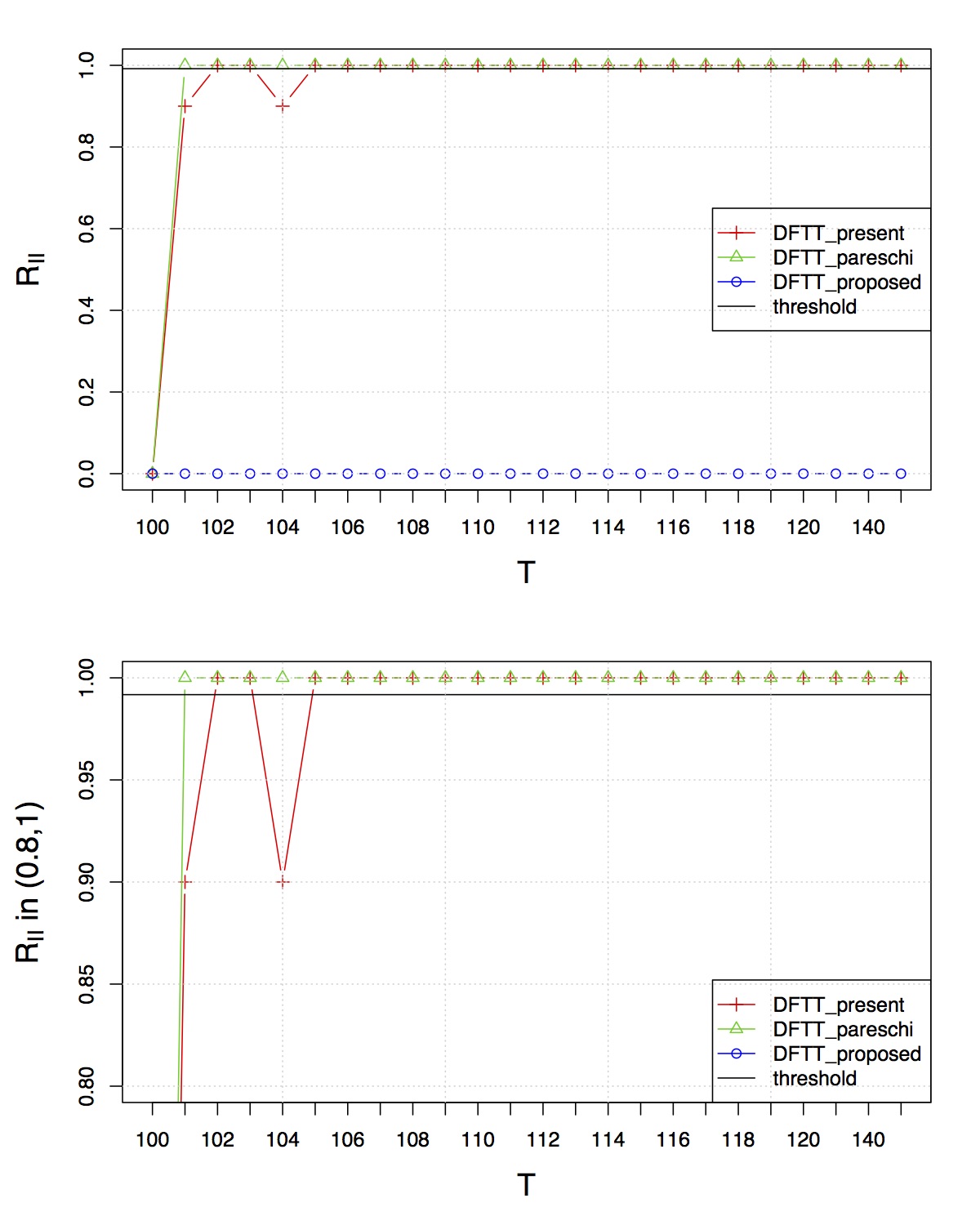}
   \caption{Passing rate $R_{\II}$ in experiment 1.
   The ``threshold'' means the lower limit of the significance
   interval defined in Eq. \eqref{eq:sig_ex1_II}
   }
   \label{fig:R_II_T}
  \end{figure}

  In this experiment, we compare the \textit{sensitivity} of \DFTpre
   and \DFTpare. \textit{Sensitivity} means \hl{a} low false negative rate
  (low probability of type I error),
  i.e., high true positive rate. Here, we compare the true positive rate
  of each test \hl{result}.
  \begin{eqnarray*}
   Sensitivity
    &:=& \mbox{low probability of type II error}\\
   &=& \mbox{low false negative rate} \\
   &=& \mbox{high true positive rate}
  \end{eqnarray*}
  Now, we define \hl{an} $nm$-length input sequence $\mathcal{X}_{n,m}$ as
  \begin{eqnarray*}
   \mathcal{X}_{n,m}
    &:=&\{x_0,x_1,x_2,\dots,x_{mn-1}\}\\
   &=&\{X^n_1,X^n_2,\dots,X^n_m\},
  \end{eqnarray*}
  where
  \[
   X^n_i= \{x_{(i-1)n},\dots,x_{in-1}\} \,\,  (i=1,2,\dots, m),
  \]
 \[
   x_k \in \{-1,1\}. \,\, (k=1,2,\dots, mn-1).
 \]
  We purposely \hl{create} non-random (periodic) sequences from the $mn$-length sequence
  $\mathcal{X}_{n,m}$ \hl{using} the method described as \hl{follows}:
  \[
    x_{k} =
    \begin{cases}
     -1 &  (k \bmod{T} =0 \,\&\, k \bmod{2T} =0 )\\
     1 &  (k \bmod{T} =0 \,\&\, k \bmod{2T} \neq 0 )
    \end{cases}.
  \]
  Therefore,
  \begin{eqnarray*}
   \mathcal{X}_{n,m}^{T}&:=&
    \{x_0,\dots,x_{T},\dots,x_{2T},\dots,x_{3T},\dots,x_{4T},\dots,x_{mn-1}\}\\
    &=&
    \{x_0,\dots,-1,\dots,1,\dots,-1,\dots,1,\dots,x_{mn-1}\}.
   \end{eqnarray*}
   We can clearly \hl{state} this sequence is \hl{a} non-random sequence.
  Therefore, if the test does not reject the $\mathcal{H}_0$ (=null
  hypothesis: ``generator is random''), \hl{then it is a} false negative (type II
  error).

  For each $T \in \{100,101,102,\dots,120,130,140,150\}$, we use $10$
  sets of \hl{an} $mn$-length
  ($nm=100,000,000$) input sequence
  $\mathcal{X}_{n,m}$ generated by the Mersenne Twister algorithm
  \cite{MT} and covert them to  non-random $mn$-length sequences
  $\mathcal{X}_{n,m}^T$.   Table~\ref{tab:ParameterNM} in Section \ref{sec:parameter} shows the
  parameters $n$ \hl{and} $m$ for each
  \hl{test}.
  In Section \ref{sec:parameter}, we explain why \hl{the} parameters $n$ \hl{and} $m$ for
  \DFTpro \hl{are} different from the other tests.
  \hl{Note} that $mn$ is the same.
  Table~\ref{tab:result_periodic}, Fig. $\ref{fig:R_I_T}$
  and Fig. $\ref{fig:R_II_T}$ show the
  \textit{\hl{passing rate}} $R_{I(\II)}$, \hl{which is defined as follows:}
  \begin{eqnarray*}
   &&
   \mbox{Passing Rate}: R_{I(\II)}\\
    &:=&
    \frac{
    \mbox{number of $\mathcal{X}^{T}_{n,m}$ passing the second-level test I
    (II)}
    }{10}.\\
   &=&
    \begin{cases}
     \mbox{True negative rate} & (\mbox{if } \Ho = \mbox{TRUE})\\
     \mbox{False negative (type II error) rate} & (\mbox{if } \Ho = \mbox{FALSE})
    \end{cases}
   \end{eqnarray*}
   Because we know that $\mathcal{X}^{T}_{n,m}$ is non-random, \hl{we know
that}  $\Ho$ =FALSE, \hl{and} the passing rate means \hl{a} false
negative rate in this experiment.
  Now, the significance levels of second-level \hl{tests} I and II are $\alpha_I\
  (=0.01)$  and
  $\alpha_{\II}\ (=0.0001)$ (defined in \eqref{eq:alpha_II}),
  respectively. Therefore, the significance
  \hl{intervals} defined in Eq. $\eqref{eq:sig_interval2}$ of $R_I$ and
  $R_{\II}$ are described as \hl{follows}:
  \begin{footnotesize}
   \begin{eqnarray}
    &&
     \left(
      1- \alpha_{I} - 2.575\sqrt{\frac{\alpha_{I}(1-\alpha_{I})}{10}},
      1- \alpha_{I} + 2.575\sqrt{\frac{\alpha_{I}(1-\alpha_{I})}{10}}
     \right) \nonumber \\
    &&\simeq (0.991,1.07)\footnotemark[1], \label{eq:sig_ex1_I}
   \end{eqnarray}
   \footnotetext[1]{
   These significance \hl{intervals} range through $1$, although $R_I$
   and $R_{\II}$ $\in [0,1]$.
   \hl{This} is because the number of sets of input \hl{sequences} in this
   experiment is $10$, and it is too small to \hl{provide a} good approximation
   (\hl{see} Section $\ref{subsec:Algorithm_DFT}$-$\ref{enum:Prop_Pass}$).
   Furthermore, $\alpha_{\II} (=0.0001)$ is very small, so the significance
   interval of $R_{\II}$ often \hl{ranges} through $1$.
   }
   \begin{eqnarray}
    &&
     \left(
      1-\alpha_{\II} -2.575\sqrt{\frac{\alpha_{\II}(1-\alpha_{\II})}{10}},
      1-\alpha_{\II} +2.575\sqrt{\frac{\alpha_{\II}(1-\alpha_{\II})}{10}}
     \right) \nonumber \\
    &&\simeq (0.9992,1.008)\footnotemark[1]. \label{eq:sig_ex1_II}
   \end{eqnarray}
  \end{footnotesize}
  Therefore, if $R_I <0.991$ or $R_{\II} <0.9992$, we can conclude that
  \hl{the} true positive rate is high, \hl{and} we can conclude that the test is
  \textit{sensitive}.

  As \hl{shown} in Table \ref{tab:result_periodic}, Fig. \ref{fig:R_I_T}
  and Fig. \ref{fig:R_II_T}, $R_I$ and $R_{\II}$
  of \DFTpro are all $0.0 \%$,
  \hl{whereas} $R_{I(\II)}$ of \DFTpre and \DFTpare are not \hl{as} low. From this
  table and the figures, we can conclude that  \DFTpro is more
  \textit{sensitive} than the other tests.


  \subsection{Experiment 2: Test results for existing pseudo-random number generators}
  \newcolumntype{P}[1]{>{\centering\arraybackslash}p{#1}}
  \begin{table}[t]
   \centering
   \caption{Test results for periodic sequences:
   passing rate $R_I$ and $R_{\II}$ for each $T$
   (\hl{red} cell means that the $R_{I(\II)}$ lies outside its significance
   interval)}
   \label{tab:result_periodic}
   \begin{tabular}{|c||P{5.6mm}|P{5.6mm}||P{5.6mm}|P{5.6mm}||P{5.6mm}|P{5.6mm}|}
    \hline
    Test  &
    \multicolumn{2}{c||}{\DFTpre}&
    \multicolumn{2}{c||}{\DFTpare}&
    \multicolumn{2}{c|}{\DFTpro}
    \\ \hline
    Passing rate  &$R_I$ &$R_{\II}$& $R_I$&$R_{\II}$&$R_I$&$R_{\II}$ \\
    \hline\hline
    $T=100$
    &\clcr 0.0&\clcr 0.0&\clcr 0.0&\clcr 0.0&\clcr 0.0&\clcr 0.0\\
    \hline
    $T=101$
    &\clcr  0.5&\clcr 0.9&\clcr 0.8&1.0&\clcr 0.0&\clcr 0.0\\
    \hline
    $T=102$
    &\clcr 0.7&1.0&1.0&1.0&\clcr 0.0&\clcr 0.0\\
    \hline
    $T=103$
    &\clcr 0.9&1.0&1.0&1.0&\clcr 0.0       &\clcr 0.0  \\
    \hline
    $T=104$
    &\clcr 0.9& \clcr 0.9 &1.0     &1.0 &\clcr 0.0     &\clcr 0.0  \\
    \hline
    $T=105$
    &\clcr 0.7&  1.0     &1.0&1.0    &\clcr 0.0     &\clcr 0.0  \\
    \hline
    $T=106$
    &  1.0    &  1.0     &1.0&1.0    & \clcr 0.0     &\clcr 0.0  \\
    \hline
    $T=107$
    &\clcr 0.9& 1.0 &\clcr 0.9 &1.0    &\clcr 0.0     &\clcr 0.0  \\
    \hline
    $T=108$
    &\clcr 0.7&  1.0    &1.0&1.0    & \clcr 0.0   &\clcr 0.0  \\
    \hline
    $T=109$
    &\clcr 0.8&  1.0    &1.0&1.0    & \clcr 0.0   &\clcr 0.0  \\
    \hline
    $T=110$
    &\clcr 0.9&  1.0    &\clcr 0.9&1.0    & \clcr 0.0   &\clcr 0.0  \\
    \hline
    $T=111$
    &\clcr 0.9&  1.0    &\clcr 0.9&1.0    & \clcr 0.0   &\clcr 0.0  \\
    \hline
    $T=112$
    &1.0 &  1.0    &1.0 &1.0    & \clcr 0.0   &\clcr 0.0  \\
    \hline
    $T=113$
    &1.0 &  1.0    &1.0 &1.0    & \clcr 0.0   &\clcr 0.0  \\
    \hline
    $T=114$
    &\clcr 0.9&  1.0    &1.0&1.0    & \clcr 0.0   &\clcr 0.0  \\
    \hline
    $T=115$
    &\clcr 0.9&  1.0    &1.0&1.0    & \clcr 0.0   &\clcr 0.0  \\
    \hline
    $T=116$
    &1.0  &  1.0    &1.0&1.0    & \clcr 0.0   &\clcr 0.0  \\
    \hline
    $T=117$
    &\clcr 0.9&  1.0    &1.0&1.0    & \clcr 0.0   &\clcr 0.0  \\
    \hline
    $T=118$
    &1.0&  1.0    &1.0&1.0    & \clcr 0.0   &\clcr 0.0  \\
    \hline
    $T=119$
    &1.0&  1.0    &1.0&1.0    & \clcr 0.0   &\clcr 0.0  \\
    \hline
    $T=120$
    &1.0&  1.0    &1.0&1.0    & \clcr 0.0   &\clcr 0.0  \\
    \hline
    $T=130$
    &1.0&  1.0    &1.0&1.0    & \clcr 0.0   &\clcr 0.0  \\
    \hline
    $T=140$
    &\clcr 0.9&  1.0    &1.0&1.0    & \clcr 0.0   &\clcr 0.0  \\
    \hline
    $T=150$
    &\clcr 0.8&  1.0    &1.0&1.0    & \clcr 0.0   &\clcr 0.0  \\
    \hline
   \end{tabular}
  \end{table}

  We use $1000$ sets of \hl{an} $mn$-length ($mn=100,000,000$)
  $\mathcal{X}_{n,m}$ input sequence generated by
   \begin{itemize}
    \setlength{\itemsep}{1mm}
    \item AES Counter Mode (AES-CTR) \cite{AES},
    \item Mersenne Twister \cite{MT},
    \item Xorshift random number generator \cite{Xorshift},
    \item Vector Stream Cipher 2.0 (VSC 2.0) \cite{VSC2},
    \item Linear congruential generator (LCG) \cite{NIST_rev},
    \item Cubic congruential generator (CCG) \cite{NIST_rev},
    \item Quadratic congruential generator I (QCG-I) \cite{NIST_rev},
    \item Quadratic congruential generator II (QCG-II) \cite{NIST_rev},
    \item Micali-Schnorr random bit generator \cite{NIST_rev}.
   \end{itemize}
   VSC 2.0 is a stream cipher based on chaos theory,
   which \hl{was} proposed by A. Iwasaki and K. Umeno \cite{VSC2}.
   We test these PRNGs using both the DFT and MS-DFT tests,
   and \hl{we} compare
   the results.
   The parameter \hl{sets} of $n$ \hl{and} $m$ are the same as
   Table~\ref{tab:ParameterNM} in Section~\ref{sec:parameter}.

\begin{figure}[t]
    \centering
    \includegraphics[width=\picWide,bb=0 0 1200 1500]{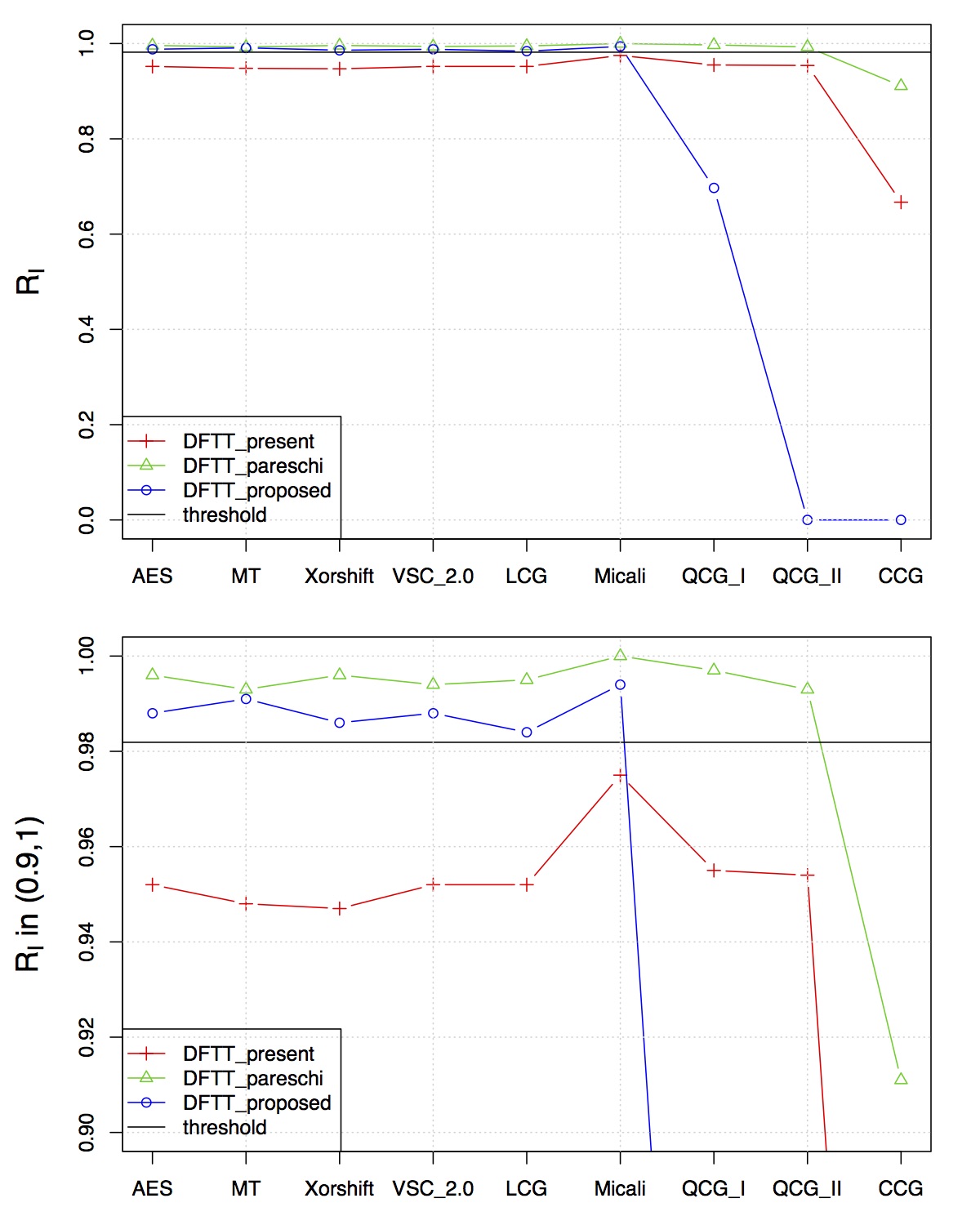}
    \caption{Passing rate $R_I$ in experiment 2.
    The ``threshold'' means the lower limit of the significance
    interval defined in Eq. \eqref{eq:sig_ex2_I}
    }
    \label{fig:R_I_PRNG}
   \end{figure}

\begin{figure}[t]
    \centering
    \includegraphics[width=\picWide,bb=0 0 1200 1500]{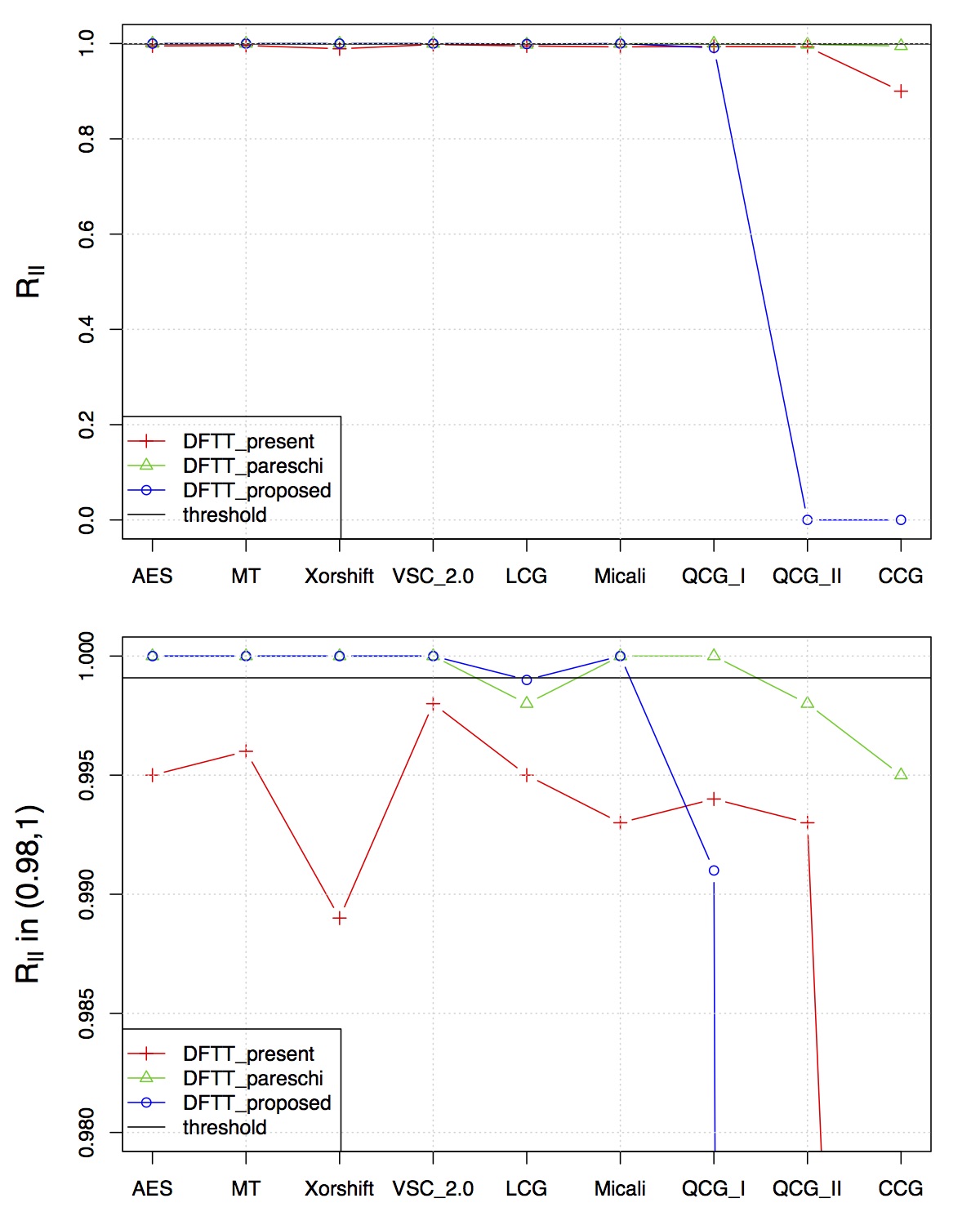}
    \caption{Passing rate $R_{\II}$ in experiment 2.
    The ``threshold'' means the lower limit of the significance
    interval defined in Eq. \eqref{eq:sig_ex2_II}
    }
    \label{fig:R_II_PRNG}
   \end{figure}

   Now, the significance levels of second-level \hl{tests} I and II are $\alpha_I:=0.01$ and
   $\alpha_{I\hspace{-.1em}I}:=0.0001$, respectively, and in this
   experiment, 1000  $mn$-length
   sequences generated by each PRNG are
   tested. Table~\ref{tab:result_PRNG}, Fig. $\ref{fig:R_I_PRNG}$
   and Fig. $\ref{fig:R_II_PRNG}$
   show the
   \textit{passing rate} $R_{I(\II)}$, \hl{defined as follows:}
   \begin{eqnarray*}
    &&
    \mbox{Passing Rate}: R_{I(\II)}\\
    &:=&
     \frac{
     \mbox{number of $\mathcal{X}_{n,m}$ passing the second-level test I
     (II)}
     }{1000}\\
    &= &
     \begin{cases}
      \mbox{True negative  rate} & (\mbox{if }  \Ho = \mbox{TRUE})\\
      \mbox{False negative (type II error) rate} & (\mbox{if } \Ho = \mbox{FALSE})
     \end{cases}
   \end{eqnarray*}
   Now, \hl{the} significance
   intervals (99\%) of passing \hl{rates} $R_I$ and $R_{\II}$
   are described as,
   \begin{footnotesize}
    \begin{eqnarray}
     &&
      \left(
       1 - \alpha_{I} - 2.575\sqrt{\frac{\alpha_{I}(1-\alpha_{I})}{1000}},
       1 - \alpha_{I} + 2.575\sqrt{\frac{\alpha_{I}(1-\alpha_{I})}{1000}}
      \right)\nonumber \\
     & \simeq& (0.9819,0.9982),
      \label{eq:sig_ex2_I}
    \end{eqnarray}
    \begin{eqnarray}
     &&
      \left(
       1 - \alpha_{\II} - 2.575\sqrt{\frac{\alpha_{\II}(1-\alpha_{\II})}{1000}},
       1 - \alpha_{\II} + 2.575\sqrt{\frac{\alpha_{\II}(1-\alpha_{\II})}{1000}}
      \right) \nonumber \\
     &\simeq& (0.9991,1.0007)\footnotemark[1],
      \label{eq:sig_ex2_II}
    \end{eqnarray}
   \end{footnotesize}
   respectively.

   In this experiment, $\Ho$ for each PRNG is defined as follows:
   \begin{itemize}
    \item $\Ho$ is TRUE (considered as random):
	  AES-CTR, Mersenne-Twister, Xorshift, VSC 2.0, LCG (Define them
	  as ``{\it good} PRNGs'').

	  Because these \hl{PRNGs} pass all the tests included in NIST
	  SP800-22 \cite{NIST_rev,VSC2}, we consider them as random in this experiment.

    \item $\Ho$ is FALSE (considered as non-random):
	  Micali-Schnorr random bit generator, QCG-I, QCG-II, CCG
	  (Define them as ``{\it bad} PRNGs'').

	  Because these \hl{PRNGs} are rejected by several tests included in NIST
	  SP800-22 \cite{NIST_rev}, we consider them as non-random in
	  this experiment.
   \end{itemize}
   Under the assumption that \hl{this} definition of $\Ho$ is appropriate,
   let us consider the \textit{sensitivity} and \textit{reliability} of
   \DFTpre, \DFTpare and \DFTpro.
   As \hl{shown} in Fig. $\ref{fig:R_II_PRNG}$, it is difficult to
   compare the \textit{reliability} from the figure. \hl{This} is because
   $R_{\II}(=0.0001)$ is very
   small, \hl{whereas} the number of sets of input \hl{sequences} is
   $1000$. Therefore, in this experiment, we focus on
   Fig. $\ref{fig:R_I_PRNG}$ and derive the conclusion of this
   experiment as follows.
   \begin{itemize}
    \setlength{\itemsep}{3mm}

    \item \textit{Reliability}; $R_I$ of ``good PRNGs'' (AES-CTR, Mersenne-Twister,
	   Xorshift, VSC 2.0, and LCG).

	  If the $R_I$ of ``good PRNGs'' \hl{lies} inside its
	  significance interval, we can conclude that the
	  \textit{reliability} of the test is \hl{sufficiently high}.

	  As \hl{shown} in Fig. $\ref{fig:R_I_PRNG}$, the $R_I$ of ``good
	  PRNGs'' of \DFTpro and \DFTpare  \hl{lies} inside its
	  significance interval, \hl{whereas} that of \DFTpre is lower than the
	  threshold. Therefore, we can conclude that the \textit{reliabilities} of
	  \DFTpare and \DFTpro are \hl{sufficiently high. Moreover,} we can
	  conclude that the \textit{reliability} of
	  \DFTpre is low.

    \item \null{\textit{Sensitivity}; the $R_I$ of ``bad PRNGs'' (Micali-
	  Schnorr random bit generator, QCG-I, QCG-II, and CCG)}.

	  If the $R_I$ of ``bad PRNGs'' \hl{lies} lower than the
	  threshold, we can conclude that the
	  \textit{sensitivity} of the test is the \hl{highest}.

	  As \hl{shown} in Fig. $\ref{fig:R_I_PRNG}$,
	  except for the Micali-Schnorr random bit generator, the $R_I$ of ``bad
	  PRNGs'' of \DFTpro are definitely lower than the
	  other tests. The $R_I$ of \DFTpre are also low, but
	  not as low as \DFTpro, and the $R_I$ of
	  \DFTpare are higher than the $R_I$ of
	  \DFTpre. Therefore, we can conclude that
	  the \textit{reliability} of \DFTpro is definitely high,
	  that of \DFTpre is high, and
	  that of \DFTpare is low.

   \end{itemize}
   These conclusions from the \hl{aforementioned experiment} are summarized in
   Table~$\rm{\ref{tab:conclusion_experiment}}$. We can conclude
   that \DFTpro is more \textit{reliable} and definitely more
   \textit{sensitive} than \DFTpre.


   \begin{table}[!t]
    \centering
    \caption{Test results for existing pseudo-random number generators:
    Passing rates $R_I$ and $R_{\II}$ of each PRNG (red cells mean that
    the $R_{I(\II)}$ lies outside its significance interval)}
    \label{tab:result_PRNG}
    \begin{tabular}{|c||P{5.6mm}|P{5.6mm}||P{5.6mm}|P{5.6mm}||P{5.6mm}|P{5.6mm}|}
     \hline
     Test  &
     \multicolumn{2}{c||}{\DFTpre}&
     \multicolumn{2}{c||}{\DFTpare}&
     \multicolumn{2}{c|}{\DFTpro}
     \\ \hline
     Passing rate  &$R_I$ &$R_{\II}$& $R_I$&$R_{\II}$&$R_I$&$R_{\II}$ \\
     \hline\hline
     AES-CTR &\clcr 0.952 &\clcr  0.995 & 0.996 &1.000 & 0.988 & 1.000\\
     \hline
     Mersenne-Twister &\clcr 0.948 &\clcr 0.996 & 0.993 & 1.000 & 0.991 &1.000\\
     \hline
     Xorshift &\clcr 0.947&\clcr 0.989 & 0.996 & 1.000& 0.986 & 1.000\\
     \hline
     VSC 2.0 &\clcr 0.952 &\clcr 0.998 & 0.994 & 1.000 &0.988 & 1.000\\
     \hline
     LCG &\clcr 0.952 &\clcr 0.995 & 0.995 &\clcr 0.998 & 0.984 &\clcr 0.999\\
     \hline
     \hline
     Micali-Schnorr &\clcr 0.975 &\clcr 0.993 & 1.000 & 1.000 & 0.994 &1.000\\
     \hline
     QCG-I &\clcr 0.955 &\clcr 0.994 & 0.997 & 1.000 &\clcr 0.697 &\clcr 0.991 \\
     \hline
     QCG-II &\clcr 0.954 &\clcr 0.993 & 0.993 &\clcr 0.998 &\clcr 0.000&\clcr 0.000 \\
     \hline
     CCG &\clcr 0.667 &\clcr 0.900 &\clcr 0.911 &\clcr 0.995&\clcr 0.000&\clcr 0.000\\
     \hline
    \end{tabular}
   \end{table}

   \begin{table}[t]
    \centering
    \caption{Summary of the conclusions derived from experiments 1 and 2}
    \label{tab:conclusion_experiment}
     \begin{tabular}{|c||c|c|c|}
      \hline
      Test&\DFTpre&\DFTpare&\DFTpro\\
      \hline\hline
      \textit{Reliability}& low & high enough & high enough \\
      \textit{Sensitivity}& high & low  & definitely high\\
      \hline
     \end{tabular}
   \end{table}


  \subsection{Appropriate selection of $n$ and $m$}
  \label{sec:parameter}
  As \hl{shown} in Table~\ref{tab:ParameterNM},
  the parameters $n$ and $m$ of \DFTpro are different from the other tests.
  NIST SP800-22 recommends $n=1,000,000$ and $m=1,000$ \cite{NIST_rev}
  (in experiments 1 and 2, we defined $n=100,000$ and $m=1,000$ for \DFTpre and
  \DFTpare \hl{to avoid excessive computation} because we need
  $10$ and $1000$ of $mn$-length sequences, respectively).
  However, as we stated in Step \ref{sec:Pvalue_DFTpro}) in
  Section \ref{sec:procedure_DFTpro}, in the procedure of \DFTpro,
  $\frac{n}{2}-1$ $P$-$value$s are generated, \hl{whereas} \DFTpre and \DFTpare
  generate $m$ $P$-$value$s.

  Pareschi \etal reported that the number of $P$-$value$s should not be too
  large because for extremely \hl{large} numbers of $P$-$value$s, the second-level
  tests always fail \cite{Pareschi2,Sackrowitz}. Pareschi \etal
  recommended that, in the case that  $n=2^{20} = 1,048,576$, for the
  frequency test included in NIST SP800-22, the number of $P$-$value$s
  should be smaller than $4795$.
  Therefore, in \DFTpro, $n$ should not
  be too large \hl{(in }\DFTpre,\hl{ $m$ should not be too large).}
  \hl{However}, as we proved in Theorem 3, $\chi_2^2$
  is the asymptotic distribution of
  $\frac{2}{n}|S_j(X)|^2$. Therefore, $n$ should be as large as possible.
  \hl{Thus}, in \DFTpro, a selection of the parameter $n$ is a trade-off between
  the error of the second-level test and the error of the distribution of
  $\frac{2}{n}|S_j(X)|^2$ (as shown in
  Table \ref{tab:trade-off}). Considering this trade-off, we
  defined the
  value of $n$ as shown in Table~\ref{tab:ParameterNM}.
  The appropriate selection of $n$ and $m$ in \DFTpro \hl{still needs} to be analyzed
  more specifically.

  \begin{table}[!t]
   \centering
   \caption{The parameter sets for each test, and the numbers of $P$-$value$s
   generated by each test}
   \label{tab:ParameterNM}
    \begin{tabular}{|c||c|c|c|}
     \hline
     Parameter&\DFTpre&\DFTpare&\DFTpro\\
     \hline\hline
     $n$&100,000&100,000&4,000\\
     $m$&1,000&1,000&25,000\\
     \hline\hline\hline
     Number of $P$-$value$s &$m=1000$&$m=1000$&$\frac{n}{2}-1=1999$ \\
     \hline
    \end{tabular}
  \end{table}

  \begin{table}[!t]
   \centering
   \caption{Trade-off in the selection of $n$ in \DFTpro}
   \label{tab:trade-off}
    \begin{tabular}{|c||c|c|}
     \hline
     $n$&small&large\\
     \hline\hline
     Second-level test&Accurate & Erroneous \\
     Distribution of $\frac{2}{n}|S_j(X)|^2$ &Erroneous&Accurate\\
     \hline
    \end{tabular}
  \end{table}


 \section{Conclusion}
 In this paper, we have considered the DFT test included in the NIST
 SP800-22 statistical test suite for random number sequences.
 The most crucial problem in the present DFT test (denoted as \DFTpre) is
 that the reference
 distribution of its test statistic is {\it not mathematically} derived
 but is \hl{rather} obtained by numerical estimation with a
 \hl{pseudo-random} number generator; the basis of the \textit{test for
 randomness} itself is based on
 a \textit{\hl{pseudo-random} number generator}. Therefore, \DFTpre cannot be
 used unless the reference
 distribution is mathematically derived.

 We {\it proved} that \hl{the} asymptotic distribution
 of the power spectrum
 is $\chi_2^2$, and based on this fact, we proposed \hl{a} new DFT test
 denoted as \DFTpro,
 whose distribution of the test statistic is {\it mathematically} derived.

 Furthermore, although appropriate selection of the parameters $n$ and $m$ for \DFTpro
 still need to be analyzed more specifically,
 \hl{the} results of testing non-random sequences and several
 \hl{pseudo-random number generators} showed that \DFTpro is more
 \textit{reliable}
 and definitely more \textit{sensitive} than \DFTpre, which is the
 current standard DFT test.


 \section*{Biography}
  \subsection*{Hiroki Okada}
   received his BSc degree
   in informatics from the Kyoto University, Japan, in Mar. 2014.
   He received his MSc degree
   in informatics from the Department of Applied Mathematics \& Physics,
   Graduate School of Informatics Kyoto University, Japan, in Mar. 2016.
   He joined KDDI Corp. in Apr. 2016.

  \subsection*{Ken Umeno}
   received his BSc degree in electronic communication
   from Waseda University, Japan, in 1990. He received his MSc and PhD
   degrees in physics from the University of Tokyo, Japan, in 1992 and
   1995, respectively. From 1998 until he joined Kyoto University as
   a Professor in 2012, he worked for Japan's Ministry of Posts and
   Telecommunications in its Communications Research Laboratory (currently
   the National Institute of Information and Communications Technology). From
   2004 to 2012, he was CEO and President of ChaosWare, Inc. He received
   the LSI IP Award in 2003 and the Telecom-System Awards in 2003 and in
   2008. He holds 46 registered Japanese patents, 23 registered United
   States patents and more than 5 international patents in the fields of
   telecommunications, security, and financial engineering. His research
   interests include ergodic theory, statistical computing, coding theory,
   chaos theory, information security, and GNSS based earthquake prediction.

\end{document}